\pdfoutput=1
\documentclass[fleqn,usenatbib]{mnras}
\usepackage{newtxtext,newtxmath}
\usepackage[T1]{fontenc}
\usepackage{ae,aecompl}
\hypersetup{draft}
\usepackage{natbib}

\usepackage{graphicx}	% Including figure files
\usepackage{amsmath}	% Advanced maths commands
\usepackage{amssymb}	% Extra maths symbols

\title[Satellites as matter tracer]{Do satellite galaxies trace matter in galaxy clusters ?}

\author[Chunxiang Wang et. al]
       {\parbox[t]{\textwidth}{
        Chunxiang Wang$^{1,2\textbf{}}$,
        Ran Li$^{1,3}$\thanks{E-mail:liran827@gmail.com},
        Liang Gao$^{1,3,4}$,
        Huanyuan Shan$^{5,6}$,
        Jean-Paul Kneib$^{5,7}$,
        Wenting Wang$^{8}$,
        Gang Chen$^{2}$,
        Martin Makler${^9}$,
        Maria~E. S.~Pereira${^9}$,
        Lin Wang$^{1,3}$,
        Marcio~A.G.~Maia${^{10}}$,
        Thomas Erben${^{6}}$}
        \vspace*{3pt} \\
   $^{1}$Key Laboratory for Computational Astrophysics, National Astronomical Observatories, Chinese Academy of Sciences, Beijing\\ 100012, China\\
   $^{2}$Tianjin Astrophysics Center, Tianjin Normal University, Tianjin 300387, China\\
   $^{3}$School of Astronomy and Space Sciences, University of Chinese Academy of Sciences,19A Yuquan Road,Beijing 100049, China \\
   $^{4}$Institute of Computational Cosmology, Department of Physics, University of Durham, Science Laboratories, South Road,
           Durham\\ DH1 3LE, UK\\
   $^{5}$Laboratoire d'astrophysique (LASTRO), Ecole Polytechnique F\'ed\'erale de Lausanne (EPFL), Observatoire de Sauverny, CH-1290 \\
           Versoix, Switzerland\\
   $^{6}$Argelander Institute for Astronomy, University of Bonn, Auf dem H\"ugel 71, D-53121 Bonn, Germany\\
   $^{7}$Aix Marseille Universit\'e, CNRS, LAM (Laboratoire d'Astrophysique de Marseille) UMR 7326, 13388, Marseille, France\\
   $^{8}$Kavli IPMU (WPI), UTIAS, The University of Tokyo, Kashiwa, Chiba 277-8583, Japan\\
   $^{9}$Centro Brasileiro de Pesquisas F\'isicas - Rua Dr. Xavier Sigaud 150, CEP 22290-180, Rio de Janeiro, RJ, Brazil\\
   $^{10}$Laborat\'orio Interinstitucional de e-Astronomia - LIneA, Rua General Jos\'e Cristino, 77, Rio de Janeiro, RJ, 20921-400, Brazil\\}

\pubyear{2018}

\begin{document}
\label{firstpage}
\pagerange{\pageref{firstpage}--\pageref{lastpage}}
\maketitle

\begin{abstract}
The spatial distribution of satellite galaxies encodes rich information of the structure and assembly history of galaxy clusters. In this paper, we select a redMaPPer cluster sample in SDSS Stripe 82 region with $0.1 \le z \le 0.33$, $20<\lambda<100$ and $P_{\rm cen}>0.7$. Using the high-quality weak lensing data from CS82 Survey, we constrain the mass profile of this sample. Then we compare directly the mass density profile with the satellite number density profile. We find that the total mass and number density profiles have the same shape, both well fitted by an NFW profile. The scale radii agree with each other within 1$\sigma$ error  ($r_{\rm s,gal}=0.34_{-0.03}^{+0.04}$Mpc vs $r_{\rm s}=0.37_{-0.10}^{+0.15}$Mpc ).

\end{abstract}


%%%%%%%%%%%%%%%%%%%%%%%%%%%%%%%%%%%%%%%%%%%%%%%%%%

%%%%%%%%%%%%%%%%% BODY OF PAPER %%%%%%%%%%%%%%%%%%

\section{Introduction}

\label{sec:intro}
The spatial distribution of satellite galaxies encodes rich information of the structure of galaxy clusters/groups. In particular, the radial number density profiles of galaxy clusters have been often used to constrain galaxy formation models \citep[e.g.][]{Gao2004,Diemand2004,Wang2014}.
 High-resolution simulations show that the distribution of subhalos is less
concentrated than the distribution of dark matter \citep{Gao2004,Springel2001,Vogelsberger2014}.
In addition, subhalos appear to have a significantly shallower radial
distribution than the observed distribution of galaxies in the inner region of clusters \citep{Gao2004}.
In hydrodynamical simulations, the galaxies can survive longer than the dark matter subhaloes.
The dissipative processes of galaxy formation make the stellar component more resistant to tidal
disruption close to cluster centres \citep{Vogelsberger2014}.
Observationally, there are lots of controversies in the literature on whether satellite galaxies unbiasedly trace
the underlying mass distribution in galaxy clusters/groups. Some studies conclude that the satellite (luminosity) distribution traces the mass distribution  \citep{ Tyson1995,Squires1996,Fischer1997,Cirimele1997,Carlberg1997,VanDerMarel2000,Rines2001,Tustin2001,Biviano_Girardi2003,Ewa2003,Kneib2003,Biviano_Girardi2003,Parker2005,Popesso2007,Sheldon2009,Wojtak_Lokas2010,Sereno2010,Bahcall2014};
 whiles some studies suggest that the spatial distribution of satellites (luminosity) are less concentrated than that of matter \citep{Rines2000,Lin2004,Hansen2005,Nagai2005,Yang2005,Budzynski2012}; still some claim luminosity distribution are actually more concentrated \citep{Koranyi1998,Carlberg2001}.

 Many of previous comparisons depend on probes of mass profiles based on real observational data, e.g. dynamical modeling methods \citep{Carlberg1997,VanDerMarel2000,Rines2000,Carlberg2001,Rines2001,Tustin2001,Biviano_Girardi2003,Ewa2003,Popesso2007}, or X-ray observation \citep{Cirimele1997,Lin2004,Budzynski2012}. Mass estimation from these probes often requires some prior assumptions on the dynamical state of galaxy clusters/groups and thus may be biased. Weak lensing method is usually considered as an unbiased probe, which is independent of the dynamical states of galaxy clusters and baryonic physics in galaxy formation. In this work, we derive mass distribution of redMaPPer clusters \citep{Rykoff2014a,Rykoff2014} using the high-quality weak lensing data from
Canada-France-Hawaii Telescope (CFHT) Stripe 82 Survey \citep[CS82;][] {Shan2014,Li2014}, and compare them directly with the satellite galaxies number density from SDSS Stripe 82 \citep{Abazajian2009,Reis2012} photometric data.

The paper is laid out as follows. In \S\ref{sec:data} we describe the data used in our work. In \S\ref{sec:nd_vs_md} 
we describe lens model and how to get the satellite galaxy number density profile of our cluster sample. 
In \S\ref{sec:result}, we show the results of this work. Finally, we summarize and discuss the implication 
of our results in \S\ref{sec:sum}. Throughout this paper, we adopt a flat $\Lambda$CDM cosmological model with  the matter density parameter $\Omega_{\rm m}=0.27$ and the Hubble parameter $H_{0}=70\mathrm{kms^{-1}}\mathrm{Mpc}^{-1}$.

\section{Data}
\label{sec:data}
\subsection{RedMaPPer cluster catalog}
\label{ssec:redmapper_cluster_catalog}
The red-sequence Matched-filter Probabilistic Percolation method \citep[redMaPPer;][]{Rykoff2014,Rykoff2014a} uses the $ugriz$ magnitudes and their errors, to group spatial concentrations of red-sequence galaxies at similar redshift into cluster.
In this paper, we use redMaPPer cluster catalog extracted from SDSS DR8, restricting to the CS82 footprint, where high quality weak lensing data is available. There are 634 clusters falling in this region. We further select our final cluster sample from these clusters using the following additional conditions:
$0.1 \le z \le 0.33$, $20<\lambda<100$ and $P_{\rm cen}>0.7$,
where $z$ is the redshift of cluster, the $\lambda$ is an optical richness
estimate indicating the number of red sequence galaxies brighter
than 0.2$L_{\ast}$ at the redshift of the cluster within a scaled
aperture which has been shown as a good mass proxy \citep{Rykoff2012}, and the
$P_{\rm cen}$ is the probability of the most likely central galaxy.
For each cluster, there are five candidate central galaxies and we
always use the position of the most likely central galaxy as the proxy of the cluster centre.
The redshift cut selects a nearly volume-limited cluster sample, the richness
cut ensures a pure and statistically meaningful sample of clusters at
all richness bins \citep{Miyatake2016} and the probability cut reduces the miscerntering problem. After applying these cuts our final sample is composed of 167 clusters.
\subsection{Lensing shear catalog}
\label{subsec:shearcatalog}
The source galaxies used in this work are taken from CS82 survey
which is an $i$-band imaging survey covering the SDSS Stripe 82
region with a median seeing $0.59^{\prime\prime}$. The CS82
fields were observed in four dithered observation with 410
seconds exposure. The limited magnitude is $i_{\rm AB} \sim 24.1$ \citep{2016JCAP...08..013B}.

The shapes of faint galaxies are measured with $lensfit$ method \citep{{Miller2007},{Miller2013}}.
Each CS82 science image is supplemented by a mask, indicating regions within which accurate photometry/shape measurements of  faint  sources cannot  be  performed. According to \citet{Erben2013}, most of science analysis
are safe with $MASK \leq 1$. We use all galaxies with weight $\omega>0$,
FITCLASS=0, $MASK \leq 1$ and $z>0$, in which $\omega$ represents an
inverse variance weight assigned to each source galaxy by $lensfit$, FITCLASS
is a star/galaxy classification provided by $lensfit$, and $z$ is the photometric redshift.

After masking out bright stars and other image artifacts, the effective survey
area reduces from 173 $deg^2$ to 129.2 $deg^2$. As the CS82 is $i$-band
imaging survey, the photometric redshifts (photo-z) are obtained by using
BPZ method \citep{{Benitez2000},{Jee2006}} and computed by \citep{Bundy2015}.
Some tests on the systematics induced by
photo-z error are shown in \citep{Li2016}.
 The total number of source galaxies in this work is 4,381,917.
\subsection{Satellite galaxy catalog}
To calculate the satellite galaxy number density of our cluster sample as described in \S\ref{ssec:redmapper_cluster_catalog}, we download a photometric galaxy catalog from SDSS Stripe 82 database by requiring the magnitude of $r$-band $($ $r_{\rm mag}$ $)$ in [17, 21] with the query provided by \citet{Reis2012}. There are 1,164,364 galaxies in the catalog. By matching this photometric catalog to the redMaPPer cluster catalog with a matching tolerance of $1.0^{\prime \prime}$, ``central galaxies"  are identified in this photometric catalog.

\section{ Theory model and method}
\label{sec:nd_vs_md}
\subsection{Lensing model}
\label{ssec:wl}
We stack lens-source pairs in 7 logarithmic radial $R$ bins from 0.03
Mpc to 1.5 Mpc. Lensing signal (excess surface density $\Delta \Sigma (R)$) is calculated by

\begin{equation}
\Delta \Sigma (R) =\overline{\Sigma(<R)}-\overline{\Sigma(R)}=\frac{\sum_{\rm ls}\omega_{\rm ls}\gamma_{\rm t}^{\rm ls}\Sigma_{\rm crit}}{\sum_{\rm ls}\omega_{\rm ls}},
\end{equation}
where
\begin{equation}
\omega_{\rm ls}=\omega_{\rm s}\Sigma_{\rm crit}^{-2},\;
\end{equation}
\begin{equation}
\Sigma_{\rm crit}=\frac{c^2}{4 \rm \pi \rm G}\frac{D_{\rm s}}{D_{\rm l}D_{\rm ls}},
\end{equation}

\noindent $\overline{\Sigma(<R)}$ is the mean surface mass
density within $R$,  $\overline{\Sigma(R)}$ is the average surface density at the projected radius $R$,  $\omega_{\rm s}$ is a weight factor introduced to account for intrinsic scatter in ellipticity and shape measurement error of each source galaxy, which is same with $\omega$ we mentioned in \S\ref{ssec:redmapper_cluster_catalog},
$\Sigma_{\rm crit}$ is the critical surface density including space
geometry information, $D_{\rm s}$ and $D_{\rm l}$ are the angular
 diameter distances of source and lens, respectively,
$D_{\rm ls}$ is the angular diameter distance between source
and lens, and $\gamma_{\rm t}$ is the tangential shear.

We apply a correction to lensing signal computed from the
multiplicative shear calibration factor $m$ as in \citet{Velander2014}:
\begin{equation}
1+K(z_{\rm l})=\frac{\sum_{\rm ls}\omega_{\rm ls}(1+m)}{\sum_{\rm ls}\omega_{\rm ls}}.
\end{equation}
Weak lensing signal can finally be obtained by:

\begin{equation}
\Delta \Sigma^{\rm cal}(R)=\frac{\Delta \Sigma (R)}{1+K(z_{\rm l})}. \
\end{equation}

\noindent Owing to large photo-z uncertainties of the source galaxies,
we remove the lens-source pairs with $z_{\rm s}-z_{\rm l}< \sigma_{z} $, where $ \sigma_{z} $ represents 1$\sigma$ error of photo-z.

The weak lensing signal is modeled as:
\begin{equation}
\Delta \Sigma(R)=\frac{M_{\rm star}}{\pi R^2}+P_{\rm cc}\Delta \Sigma_{\rm NFW}(R)+(1-P_{\rm cc})\Delta \Sigma_{\rm NFW}^{\rm off}(R),\
\end{equation}
\noindent  where the first term represents the contribution of the stellar mass of the central galaxy, the second and the third terms represent the perfectly centered and miscentered component of dark matter halos (and also the diffused baryonic matter like hot gas), respectively.

 We model the central galaxy as a point mass following \citet{Leauthaud2012}
and fix $M_{\rm star}$ to the average mass of central galaxies.
Stellar masses are estimated for member galaxies in the redMaPPer
catalog using the Bayesian spectral energy distribution (SED)
modeling code ISEDFIT \citep{Moustakas2013}. $P_{\rm cc}$ and $(1-P_{\rm cc})$ are weights
for the centered and miscentering part of the dark matter halo surface
mass density, respectively.

Dark matter density profile is described by the \citet[][hereafter NFW]{Navarro1997} profile:
\begin{equation}
  \rho(r) \propto \frac{1}{(r/r_{\rm s})(1+r/r_{\rm s})^2},
  \label{equ:nfw_model}
\end{equation}

\noindent  where $r_{\rm s}$ is the scale radius which is commonly quantified in terms of the concentration parameter $C_{200}=R_{200}/r_{\rm s}$, where $R_{200}$ is the virial radius enclosing the virial mass $M_{200}=(800/3)\pi R_{200}^{3}\rho_{\rm c}$, where $\rho_{\rm c}$ is the critical density of the universe at the redshift of the halo.

 By integrating the three-dimensional density profile along the line of sight, we can get the projected surface density $\Sigma_{\rm NFW}(R)$ which is a function of the projection radius $R$:
\begin{equation}
  \Sigma_{\rm NFW}(R)=\int_{0}^{\infty} \rho\left(\sqrt{R^2+z^2}\right)dz.\
\end{equation}
 Integrating $\Sigma_{\rm NFW}(R)$ from 0 to $R$, we can get the the mean surface density within $R$,  $\overline{\Sigma_{\rm NFW}(<R)}$:
\begin{equation}
 \overline{\Sigma_{\rm NFW}(<R)}=\frac{2}{R^2}\int_{0}^{R}R'\Sigma_{\rm NFW}(R')dR',
\end{equation}
\noindent here $\rho$ is the NFW density profile.

There are possibilities that BCG may be misidentified in the cluster
catalog, we also including a $``miscentering"$ term. If the central
galaxy is offset from the halo center by a distance $R_{\rm mc}$, the mass surface density will be changed as follow:

\begin{equation}
  \Sigma_{\rm NFW}(R|R_{\rm mc})=\ \int_{0}^{2\pi} d\theta \Sigma_{\rm NFW}\left(\sqrt{R^{2}+R_{\rm mc}^{2}+2RR_{\rm mc}cos(\theta)}\right).\
\end{equation}

The distribution of miscentering can be described by a 2D Gaussian distribution:
\begin{equation}
    P(R_{\rm mc})=\frac{R_{\rm mc}}{\sigma^2_{\rm off}}exp\left(-\frac{1}{2}(\frac{R_{\rm mc}}{\sigma_{\rm off}})^2\right).
\label{equ:PRs}
\end{equation}

 In the fitting model there are four free parameters, $M_{200}$, $C_{200}$, $\sigma_{\rm off}$ and $P_{\rm cc}$.
Due to the strong degeneracy between  $\sigma_{\rm off}$ and $P_{\rm cc}$, our data are not good enough to fit $\sigma_{\rm off}$ and $P_{\rm cc}$ well synchronously (see the results in the APPENDIX~\ref{sec:appendix_a}). We assume that the position of one of the five central galaxy candidates is true center of the galaxy cluster, so we fix $\sigma_{\rm off}$ and $P_{\rm cc}$ in following way.

\begin{figure}
\begin{center}
\hspace{1.cm}
\resizebox{9.0cm}{!}
{\includegraphics{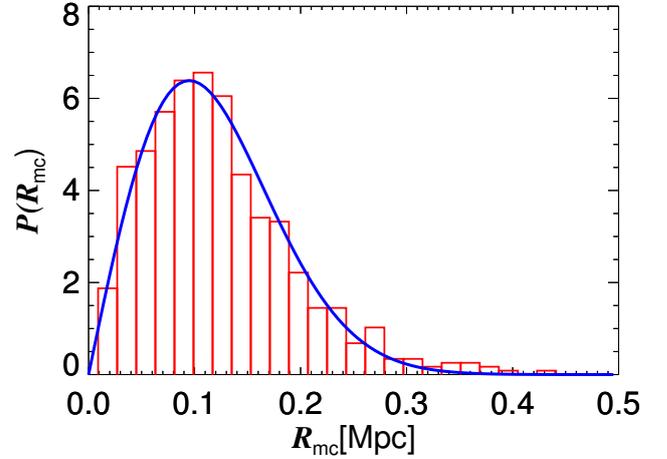}}\\%
\caption{The distribution of miscentering.
We take the projected distance between the most likely central galaxy and the 4 remaining central candidate galaxies as $R_{\rm mc}$.
The red histogram shows the distribution of  $R_{\rm mc}$. The blue line is the best fit curve of the distribution of $R_{\rm mc}$.}
\label{fig:sigma_s_fit}
\end{center}
\end{figure}

First, we fix $P_{\rm cc}=0.95$ to the average of $P_{\rm cen}$ of 167 clusters sample we finally select.
Second, we fit the distribution of the candidates of the central galaxy to obtain $\sigma_{\rm off}$.
There are five candidates of the central galaxy. We calculate
the distribution of the projected distance between the most
likely central galaxy and the 4 remaining central candidate
galaxies, and fit this distribution with Equation~(\ref{equ:PRs}).
As shown in Figure \ref{fig:sigma_s_fit}, the red histogram shows
the distribution of miscentering and the blue solid line represents the best fit curve. The best fit effective scale length is $\sigma_{\rm off}=(0.095 \pm 0.002)$Mpc.

As a comparison, we also show the four free parameters model fitting results in APPENDIX~\ref{sec:appendix_a}.

%Figure1

 Substitute Equation~(\ref{equ:PRs}) into following Equation~(\ref{equ:sigma_off}), we can obtain the resulting mean surface mass profile for the miscentered clusters.
\begin{equation}
    \Sigma_{\rm NFW}^{\rm off}(R)=\ \int dR_{\rm mc} P(R_{\rm mc}) \Sigma_{\rm NFW}(R|R_{\rm mc}).
    \label{equ:sigma_off}
\end{equation}

There are two free parameters $M_{200}$ and $C_{200}$ in our lensing fitting model.

%Figure2
\begin{figure}
\begin{center}
\hspace{1.cm}
\resizebox{9.0cm}{!}
{\includegraphics{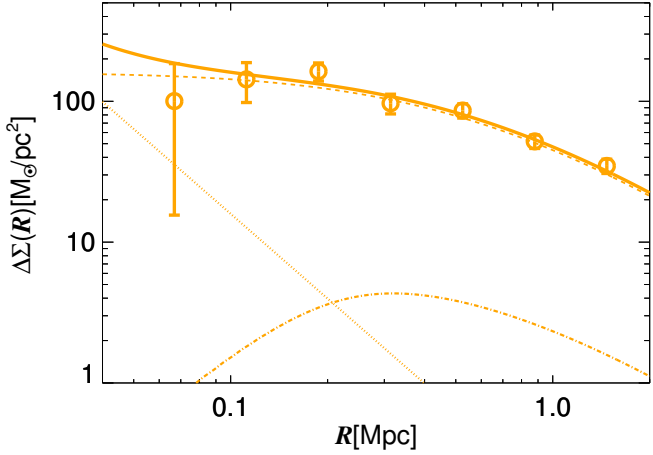}}\\%
\caption{Best-fit model for weak lensing of 167 clusters.
The orange circles represent the excess surface mass
density $\Delta \Sigma(R)$ of cluster sample.
Errors bars reflect the $68\%$ confidence intervals obtained using
bootstrapping. The solid line shows the best-fit model.
The dashed line is the centered dark matter halo term.
The dot-dashed line is the miscentering dark matter halo
term and the dotted line corresponds to the stellar mass
contribution from the central galaxy. The corresponding best-fit
parameters are listed in Table \ref{tab:167wl}.}
\label{fig:167wl}
\end{center}
\end{figure}
%
%

% Table 1
\begin{table*}
\begin{center}
\caption{ Best fit parameters of the mass profile from fitting the weak lensing data.}
\begin{tabular}{c|c|c|c|c|c|c}
 \hline  \hline
 &&&&&\\
    $M_{200}/ \rm 10^{14} \rm M_{\odot}$
& $C_{200}$
& $M_{\rm star}/ \rm 10^{11} \rm M_{\odot}$
& $r_{\rm s}/ \rm Mpc$
& $\chi^{2}/dof$  \\

 &&&&&\\
  \hline
 &&&&& \\
    $2.06_{-0.41}^{+0.61}$
 & $2.80_{-0.67}^{+0.81}$
 & $4.99$
 & $0.37_{-0.10}^{+0.15}$
 &$4.15/5$ \\
 &&&&& \\
 \hline
\end{tabular}
\label{tab:167wl}
\end{center}
\end{table*}

\subsection{Satellite number density}
\label{ssec:nd}
For each central galaxy, we count the number of galaxies in $r$-band magnitude range $17<r_{\rm mag}<21$ and not brighter than the central galaxy in different projected radial bins. These galaxies contain
satellites and galaxies in the background or foreground.

To compare directly with the weak lensing measurement,
we calculate $\Delta \Sigma_{\rm g} (R)$ instead of $\Sigma_{\rm g} (R)$,
\begin{equation}
\Delta \Sigma_{\rm g} (R) =\overline{\Sigma_{\rm g}(<R)}-\overline{\Sigma_{\rm g}(R)},
\end{equation}
\noindent where $\overline{\Sigma_{\rm g}(<R)}$ represents galaxy
surface number density within $R$, and $\overline{\Sigma_{\rm g}(R)}$ is the average galaxy surface number density at the projected radius $R$ and  each of them contains the background galaxy density.
So naturally the background is cancelled when we stack a lot of clusters. We calculate $\Delta \Sigma_{\rm g}(R)$ for each individual
cluster and average over the whole sample.

%Figure3
\begin{figure}
\begin{center}
\hspace{1.cm}
\resizebox{9.0cm}{!}
{\includegraphics{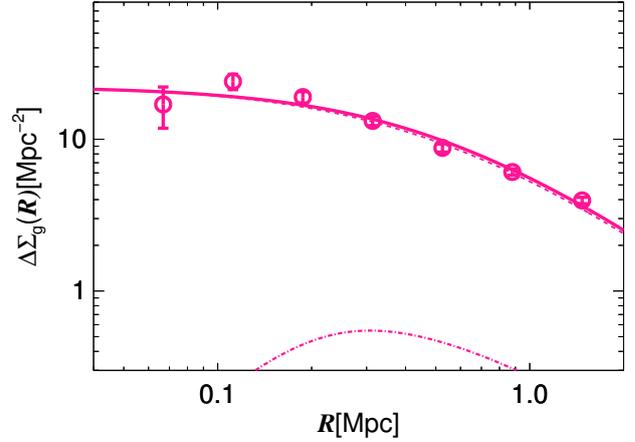}}\\%
\caption{Best-fit model for galaxy number density of
167 clusters.  The deep pink circles with errors bars represent the
excess surface number density $\Delta \Sigma_{\rm g}(R)$
of cluster sample. The solid line represents the best-fit model.
The dashed line is the centered term and the dot-dashed line is the miscentering term.
The corresponding best-fit parameters are listed in Table \ref{tab:167num}.}
\label{fig:167num}
\end{center}
\end{figure}

We assume the number density of galaxies also follow a NFW form as:
\begin{equation}
      N(r) =\frac{N_{0}}{(r/r_{\rm s,gal})(1+r/r_{\rm s,gal})^{2}}.
\end{equation}
\noindent The satellite galaxy surface number density fitting model includes
the two components:
\begin{equation}
\Delta \Sigma_{\rm g}(R)=P_{\rm cc}\Delta \Sigma_{\rm g}^{\rm cen}(R)+(1-P_{\rm cc})\Delta \Sigma_{\rm g}^{\rm off}(R).
\end{equation}
The two terms on the right side of the equation represent
centered and miscentering NFW profile, respectively.
$N_{\rm 0}$, $r_{\rm s}$ are free parameters in our fitting.
 Owing to the same center we used both in weak lensing signal calculation and satellite galaxy count, the satellite number density profile shares the same $\sigma_{\rm off}$ and $P_{\rm cc}$ with density of mass. We fix $\sigma_{\rm off}=0.095 \rm Mpc$, $P_{\rm cc}= 0.95$.

\section{ Results}
\label{sec:result}

With the Markov Chain Monte Carlo (MCMC) technique,
we can fit the weak lensing signal and the satellite galaxy number
density to get the posterior distribution of the free parameters.

In Figure \ref{fig:167wl},  we show the stacked lensing signal of our cluster sample.
The orange circles with errors bars represent weak lensing signal and
errors bars reflect the $68\%$ confidence intervals obtained by
bootstrapping. The bold solid line shows
the best-fit model, the dashed line is the centered dark
matter halo term, the dot-dashed line is the miscentring
dark matter halo term and the dotted line corresponds to the
stellar mass contribution from central galaxy.
The best fit parameters are listed in Table \ref{tab:167wl}.
We obtain a halo mass $M_{200}=2.06_{-0.41}^{+0.61} \times \rm 10^{14} \rm M_{\odot}$
that is consistent with the halo mass fitting result in \citet{Miyatake2016}, as well as the
halo mass estimated by mass-richness relation in \citet{Melchior2016} and \citet{Shan2017} within $1\sigma$ error.
The fitted scale radius is $r_{\rm s}=0.37_{-0.10}^{+0.15} \rm Mpc$. The concentration parameter obtained here is $C_{200}=2.80_{-0.67}^{+0.81}$.
 To compare our measurements with the three-dimensional (3D) N-body simulation results directly, we correct the $C_{200}$ with the 3D correction in \citet{Giocoli2012}:
\begin{equation}
C_{\rm 2D}(M) = C_{\rm 3D}(M) \times 1.630M^{-0.018},
\end{equation}

\noindent  and rescale the concentration parameter to $z=0$ with the redshift dependence in \citet{Klypin2016}.
We get the corrected concentration parameter $C_{\rm 200,3D}=3.62_{-0.88}^{+1.07}$, which is consistent with the prediction from cosmological simulations provided by \citet{Klypin2016} within $1\sigma$ error.

In Figure \ref{fig:167num}, we show the excess surface number density of satellite galaxy of our cluster sample.
The deep pink circles with errors bars are the satellite galaxy excess number surface density. The solid line represents
the best-fit model. The dashed line is the centered term and the dot-dashed line is the miscentering term.
Fitting results of excess surface number density are listed in Table \ref{tab:167num}.

\begin{center}
\begin{figure*}
\includegraphics[width=0.8\linewidth]{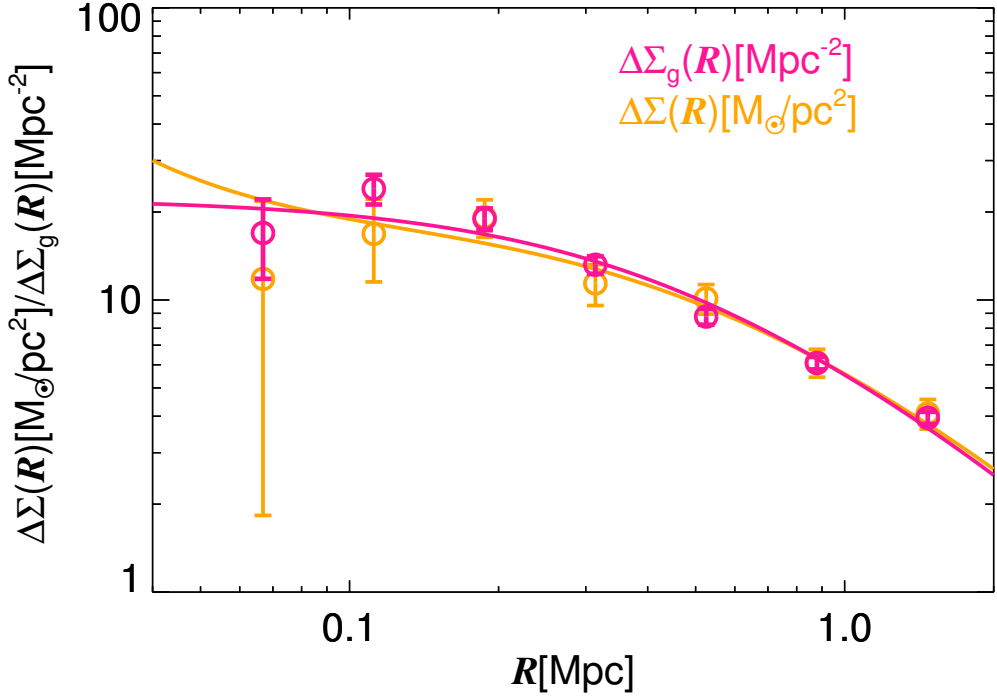}
\caption{Comparison between mass distribution and galaxy number
density profile. The deep pink circles with errors bars correspond to the satellite
 galaxy excess surface number density $\Delta \Sigma_{\rm g}(R)$ and
 the deep pink solid line represents the best-fit model.
 Orange circles represent excess mass surface density $\Delta \Sigma(R)$
 and the solid orange line represent the best-fit model.
 Errors bars reflect 1$\sigma$ uncertainties.
  Here we divided 8.5 into $\Delta \Sigma(R)$ to get a similar
  amplitude with $\Delta \Sigma_{\rm g}(R)$. Satellite galaxies are selected by
$17<r_{\rm mag}<21$. }
\label{fig:167nd_vs_md}
\end{figure*}
\end{center}

We compare the satellite galaxy excess surface number density  $\Delta \Sigma_{\rm g}(R)$ with
the mass excess surface density $\Delta \Sigma(R)$ directly in Figure \ref{fig:167nd_vs_md}.
To compare their profiles intuitively, we divide 8.5 into $\Delta \Sigma(R)$ to obtain a similar amplitude with $\Delta \Sigma_{\rm g}(R)$.
As shown in Figure \ref{fig:167nd_vs_md},
they have similar distribution.
We find that the fitted scale radius with satellite galaxy excess
surface number density $r_{\rm s,gal}=  0.34_{-0.03}^{+0.04} \rm Mpc$ ($C_{\rm g}=3.03 \pm 0.30$)
 is consistent with the scale radius $r_{\rm s}=0.37_{-0.10}^{+0.15} \rm Mpc$ ($C_{200}=2.80_{-0.67}^{+0.81}$)
 fitted with weak lensing signal within $1\sigma$ error showing that the satellite
 galaxy number density profile traces mass distribution closely in the galaxy clusters.

\begin{figure}
\begin{center}
\hspace{1.cm}
\resizebox{9.0cm}{!}
{\includegraphics{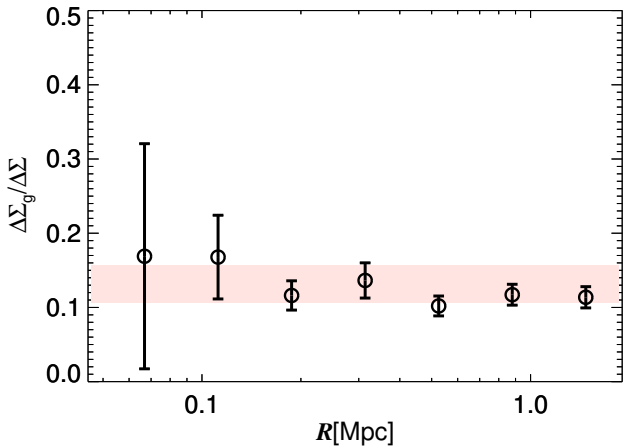}}\\%
\caption{The distribution of number-to-mass ratio. Errors bars reflect the 1$\sigma$ uncertainties. The shaded region is standard errors of the number-to-mass ratio.}
\label{fig:167ratio}
\end{center}
\end{figure}

 In some previous studies, the generalized NFW or the Einasto parametric profile model is also used to fit the mass density or satellite galaxies number density profile \citep{2016ApJ...825...39M,Ewa2003}. In this paper, only the NFW profile model is adopted. Thus we also compare these two profiles in a non-parametric way without any model dependence. In Figure~\ref{fig:167ratio}, we show the distribution of number-to-mass ratio with the projected radius $R$. Errors bars represent the 1$\sigma$ uncertainties. The shaded region is standard errors of the number-to-mass ratio. The number-to-mass ratio is nearly a constant within $1\sigma$ error. Note that the number-to-mass ratio is still nearly a constant when projected distances are scaled by virial radii from the mass-richness scaling relation in \citet{Simet2017}.

 %Table 2
\begin{table}
\begin{center}
\caption{ Best fit parameters of the galaxy density profile.}
\begin{tabular}{lc|clc|}
 \hline  \hline
 &&&&\\
  $N_{0}/\rm Mpc^{-3}$
 & $r_{\rm s,gal}/\rm Mpc$
 & $C_{\rm g}$
 &$\chi^{2}/dof$\\
 &&&&\\
  \hline
 &&&&\\
    $68.17_{-11.26}^{+13.60}  $
 & $ 0.34_{-0.03}^{+0.04} $
 & $3.03 \pm 0.30$
 & $12.27/5$\\
 &&&& \\
 \hline
\end{tabular}
\label{tab:167num}
\end{center}
\end{table}
%

%Figure4
%

%Figure Question_1_5

\section{Summary}
\label{sec:sum}
In this short paper, we perform a comparison between the satellite number density profile and mass profile of redMaPPer clusters. For the mass profile, we select a sample of 167 redMaPPer clusters in the CS82 area with
$20<\lambda<100$, $0.1 \le z \le 0.33$ and $P_{\rm cen}>0.7$ and calculate
the stacked weak lensing signal around them to obtain the mass distribution from 0.03 Mpc to 1.5 Mpc. We extract the satellite galaxies in the same cluster sample using SDSS Stripe 82 photometric data in the $r$-band magnitude range $17<r_{\rm mag}<21$.
Comparing the excess surface mass density with the satellite galaxy number density, we find that they agree with each other well and can both be fitted with the NFW profile. The best-fit scale radius $r_{\rm s}$ and concentration parameter $C$ of these two profiles are consistent with each other within $1\sigma$ error, thus we can conclude that the satellite galaxy number density is an unbiased tracers of mass distribution in galaxy clusters. Our conclusion is consistent with some similar studies using observational data based on dynamical methods \citep[e.g.][]{Carlberg1997,VanDerMarel2000,Biviano_Girardi2003} or based on the other methods \citep[e.g.][]{Cirimele1997,Parker2005,Sereno2010}.

\section*{Acknowledgements}
We are indebted to the referee the thoughtful comments and
insightful suggestions that improved this paper greatly. 
Based on observations obtained withMegaPrime/MegaCam, a joint
project of CFHT and CEA/DAPNIA, at the CFHT, which is operated
by the National Research Council (NRC) of Canada, the
Institut National des Science de l'Univers of the Centre National
de la Recherche Scientifique (CNRS) of France and the University
of Hawaii. The Brazilian partnership on CFHT is managed by the
Laboratrio Nacional de Astronomia (LNA). This work made use
of the CHE cluster, managed and funded by ICRA/CBPF/MCTI,
with financial support from FINEP and FAPERJ. We thank the support
of the Laboratrio Interinstitucional de e-Astronomia (LIneA).
We thank the CFHTLenS team for their pipeline development and
verification upon which much of this surveys pipeline was built.

We acknowledge support from the National Key Program for Science and Technology Research and Development (2017YFB0203300).
RL acknowledges NSFC grant (Nos. 11773032, 11333001),
support from the Youth Innovation Promotion Association of CAS, Youth Science
funding of NAOC and Nebula Talent Program of NAOC.
LG acknowledges support from the NSFC grant (Nos. 11133003, 11425312),
and a Newton Advanced Fellowship, as well as the hospitality of the Institute
for Computational Cosmology at Durham University. HYS and JPK acknowledge
support from the ERC advanced grant LIDA. CXW and GC acknowledge
NSFC grant No. 10903006,  and support from the Middle-aged and Young Key
Innovative Talents Program for Universities in Tianjin.
M.~Makler is partially supported by CNPq and FAPERJ.
Fora Temer. T.~Erben supported by the Deutsche Forschungsgemeinschaft
in the framework of the TR33 `The Dark Universe'.



\bibliographystyle{mnras}
\bibliography{wang} % if your bibtex file is called example.bib


%%%%%%%%%%%%%%%%%%%%%%%%%%%%%%%%%%%%%%%%%%%%%%%%%%

%%%%%%%%%%%%%%%%%%%%%%%%%%%%%%%%%%%%%%%%%%%%%%%%%%
\appendix
\section{Four free parameters model}
\label{sec:appendix_a}

\begin{figure}
\begin{center}
\hspace{1.cm}
\resizebox{9.0cm}{!}
{\includegraphics{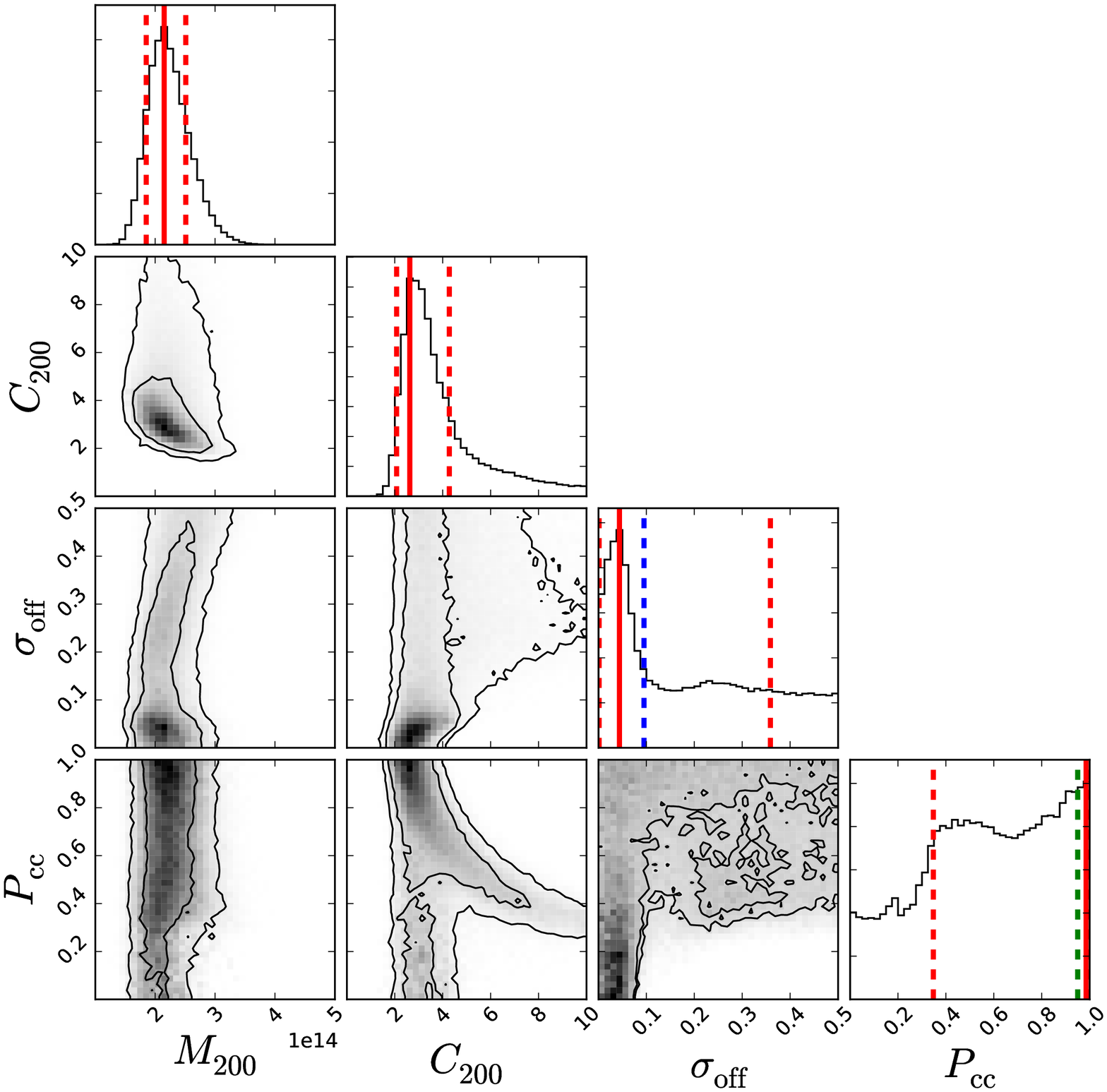}}\\%
\caption{ The 68 and 95 per cent confidence intervals for the four free parameters, $M_{200}$, $C_{200}$, $\sigma_{\rm off}$ and $P_{\rm cc}$. $M_{200}$ and $\sigma_{\rm off}$ are in units of $M_{\odot}$ and $\rm Mpc$, respectively. The last panel in each row shows the marginalized posterior distribution and the red solid lines represent the best fit parameters. The red dashed lines are the $1\sigma$ error of $\sigma_{\rm off}$ and $P_{\rm cc}$. The blue dashed line is $\sigma_{\rm off}=(0.095 \pm 0.002)$Mpc and the green dashed line is $P_{\rm cc}=0.95$ which are the values we used in our two-parameter model.}
\label{fig:Question_1}
\end{center}
\end{figure}

 In the lensing model, we can also treat $\sigma_{\rm off}$ and $P_{\rm cc}$ as free parameters. Thus, we have four free parameters in the fitting model, $M_{200}$, $C_{200}$,  $\sigma_{\rm off}$ and $P_{\rm cc}$. We show the 68 and 95 percent confidence intervals for the four free parameters in Figure~\ref{fig:Question_1}. The last panel in each row shows the marginalized posterior distribution and the red solid lines represent  the best
fitting parameters. The red dashed lines are the $1\sigma$ error of $\sigma_{\rm off}$ and $P_{\rm cc}$. The blue dashed lines represents the value of $\sigma_{\rm off}$ and $P_{\rm cc}$ in our two-parameters model.

For weak lensing data fitting, we obtain a halo mass $M_{200}=2.15_{-0.32}^{+0.38} \times \rm 10^{14} \rm M_{\odot}$ and concentration parameter $C_{200}=2.63_{-0.61}^{+1.80}$ which are consistent with our two-parameters model results in Section~\ref{ssec:wl} within $1\sigma$ error. The best-fit results are listed in Table~\ref{tab:167wl_4freepar}.

\begin{figure}
\begin{center}
\hspace{1.cm}
\resizebox{9.0cm}{!}
{\includegraphics{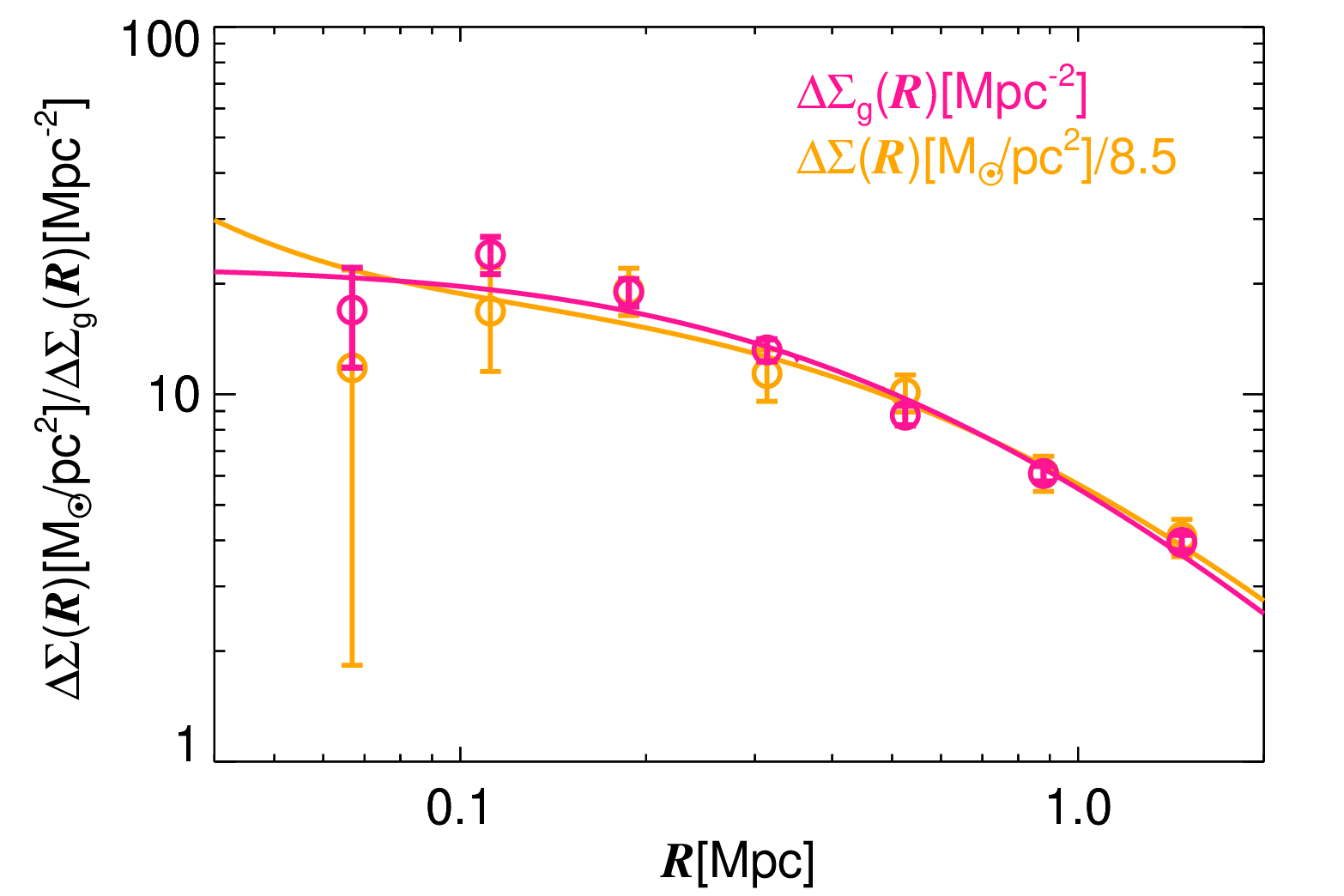}}\\%
\caption{Similar figure to Figure~\ref{fig:167nd_vs_md}, but with 4 free parameters.}
\label{fig:167nd_vs_md_4freepar}
\end{center}
\end{figure}

As the satellite number density shares the same $\sigma_{\rm off}$ and $P_{\rm cc}$ with density of mass, we thus fix the two parameters to the best-fit value from weak lensing data for the satellite number density fitting. We show the best-fitting model of satellite number density in Table~\ref{tab:167num_4freepar}. Again, the best-fit scale radius from the galaxy density profile agrees with that from the lensing data (see Figure~\ref{fig:167nd_vs_md_4freepar}). \\

%Figure A1

% Table
\begin{table*}
\begin{center}
\caption{ Best fit parameters of the mass profile from fitting the weak lensing data}
\begin{tabular}{c|c|c|c|c|c|clclc}
 \hline  \hline
 &&&&&&&\\
    $M_{200}/ \rm 10^{14} \rm M_{\odot}$
& $C_{200}$
& $M_{\rm star}/ \rm 10^{11} \rm M_{\odot}$
& $\sigma_{\rm off}/ \rm Mpc$
& $P_{\rm cc}$
& $r_{\rm s}/ \rm Mpc$
& $\chi^{2}/dof$  \\

 &&&&&&&\\
  \hline
 &&&&&&& \\
    $2.15_{-0.32}^{+0.38} $
 & $2.63_{-0.61}^{+1.80}$
 & $4.99$
 & $0.04_{-0.04}^{+0.32}$
 & $0.99_{-0.64}^{+0.01}$
 & $0.32_{-0.14}^{+0.10}$
 &$3.967/3$ \\
 &&&&&&& \\
 \hline
\end{tabular}
\label{tab:167wl_4freepar}
\end{center}
\end{table*}
%

%Table Question_1_2

\begin{table}
\begin{center}
\caption{ Best fit parameters of the galaxy density profile. }
\begin{tabular}{lc|clc|}
 \hline  \hline
 &&&&\\
  $N_{0}/\rm Mpc^{-3}$
 & $r_{\rm s,gal}/\rm Mpc$
 & $C_{\rm g}$
 &$\chi^{2}/dof$\\
 &&&&\\
  \hline
 &&&&\\
    $64.353_{-10.47}^{+12.41}  $
 & $ 0.35_{-0.03}^{+0.04} $
 & $2.98 \pm 0.17$
 & $11.342/5$\\
 &&&& \\
 \hline
\end{tabular}
\label{tab:167num_4freepar}
\end{center}
\end{table}
%

%Figure A2
%

% Don't change these lines
\bsp	% typesetting comment
\label{lastpage}
\end{document}